\documentclass[preprint,
amsmath,amssymb,
aps,
]{revtex4-2}

\usepackage{graphicx}
\usepackage{amssymb}
\usepackage{amsfonts}
\usepackage{amsmath}
\usepackage{amsbsy}
\usepackage{natbib}
\usepackage{comment}

\usepackage{color}
\usepackage{float}

\newcommand{\ii}{\mathrm{i}}
\newcommand{\mr}{\mathrm}
\newcommand{\ee}{\mathrm{e}}
\newcommand{\dd}{\mathrm{d}}
\renewcommand{\vec}[1]{\boldsymbol{\mathrm{#1}}}
\newcommand{\bcorr}{\color{black}}

\graphicspath{{figures/}}

\begin{document}

\title{Magnon dynamics in parity-time symmetric, dipolarly coupled waveguides and magnonic crystals}
\author{Xi-guang Wang$^{1}$, Dominik Schulz$^2$, Guang-hua Guo$^{1}$, Jamal Berakdar$^2$}

\address{$^1$ School of Physics and Electronics, Central South University, Changsha 410083, China \\
	$^2$ Institut f\"ur Physik, Martin-Luther Universit\"at Halle-Wittenberg, 06099 Halle/Saale, Germany}

\date{\today}

\begin{abstract}
	We consider the magnonic properties of two dipolarly coupled magnetic stripes, both deposited on a normal conductive substrate with strong spin-orbit coupling. A charge current in the substrate acts on the adjacent magnets with spin-orbit torques, which result in magnonic damping or antidamping of the spin waves, and hence a gain-loss coupling of the two magnetic stripes. The whole setup is demonstrated to exhibit features typical for parity-time (PT) symmetric systems. Phenomena are demonstrated that can be functionalized in magnonic devices{\bcorr, including reconfigurable magnonic diodes and logic devices}. 
Alternative stripes designs and PT-symmetric, periodic, coupled magnonic textures are studied. Analytical and full numerical analysis identify the conditions for 
 the appearance of exceptional points (EPs), where magnonic gain and loss are balanced and evidence nonreciprocal magnon propagation and enhanced magnon excitation around EPs. Furthermore, the dipolar coupling is shown to bring in a wave vector-dependent PT-symmetric behavior. 
Proposing and simulating a PT-symmetric magnonic crystal, we show how EPs and hence associated phenomena can be steered to a particular wave vector in a gaped spectrum via material design. 
The phenomena offer additional tools for magnonic-based communication and computational devices. 
\end{abstract}

\maketitle

\section{Introduction}
Magnons are the quanta of low-energy collective (spin wave, SW) excitations in a magnetically ordered material.
 Magnonic devices utilize SW signals to transfer and/or process information at a relatively low energy cost, high operation frequency, and without Joule heating.\cite{Chumak2015,Kruglyak2010,Serga2010} Proposals and realizations of such devices include, for example, magnonic logic gates, phase shifters, and magnon transistors.\cite{PhysRevX.5.041049,Mahmoud2020,xiguang2018,Khitun2010,Vogt2014,Lee52008,Chumak,Balynsky2017} 
 A good magnonic device performance goes along with efficient and versatile control of the magnon propagation in the SW channels.
 In this context, coupled magnonic waveguides are essential building blocks as they allow steering and transmitting the SW signals to various units in a magnonic circuit.\cite{Wang2018,Sadovnikov2015hm,Wangq2020,Sadovnikov2018,Ren2019,Sadovnikov2017,FanYabin2020,Wangxi2019} 
 A key enabling feature is the magnetic coupling between the waveguides, which renders possible the operation of magnonic directional coupler, power splitter, multiplexer, half-adder, and others.\cite{Wang2018,Sadovnikov2015hm,Wangq2020,Sadovnikov2018,Ren2019,Sadovnikov2017,FanYabin2020,Wangxi2019} 
 External tools to trigger and act on the SW signal include magnetic and electric fields, or spin-transfer and spin-orbit torques. On the other hand, the limiting factors for applications include the relatively long response time of magnetic dynamics, the necessity for a strong magnetoelectric coupling or large charge-current densities if electric fields are to be used.
 Thus, alternative design proposals are helpful and can shed light on different physical aspects of coupled SW excitations. 
 
 {\bcorr In recent years, non-Hermitian PT-symmetric systems have been discussed in quantum mechanics\cite{PhysRevLett.80.5243,PhysRevLett.89.270401,Bender2007}, optics\cite{Feng2014,Ganainy2018,Ruter2010,Regensburger2012,Miri2019}, acoustics\cite{Zhu2014,Fleury2015}, electronics\cite{Assawaworrarit2017,Chen2018}, and magnonics\cite{Lee2015,Zhang2017,Yang2018,Galda2019,Liu2019, Wangxinc2020, PhysRevApplied034050, wangxiapl2020,Sui2022}. For such systems, it is possible to achieve an exceptional point (EP) where the PT-symmetry phase breaking occurs.\cite{Feng2014, Ganainy2018, Ruter2010,Miri2019,PhysRevApplied034050} Near the EP, the PT-symmetric systems possess several unusual properties, which are widely discussed for various promising applications.\cite{Hodaei2016,Wiersig2020,Zhang2018,Lai2019,PhysRevLett.112.203901} For example, by introducing balanced gain and loss in coupled magnetic layers, magnonic PT-symmetry was generated. Gain (loss) here means antidamping (damping) of magnonic excitations. Varying the  ratio between  gain and loss, the coupled magnonic system can be tuned to the PT-symmetry broken phase at the EP, where two or more magnon eigenmodes coalesce simultaneously. In this process, the orthogonality between different eigenmodes is gradually weakened, leading to a non-reciprocal magnon propagation. Injecting magnons in different magnetic layers, the magnon profile is then not reciprocally switched.\cite{ Wangxinc2020} In the vicinity of the EP,  the magnon excitation efficiency and the response to external disturbance are substantially  enhanced.\cite{Wangxinc2020,Yu2020,PhysRevApplied034050} Above the EP, an abrupt spin reversal is observed. Besides, by varying appropriately some external parameters (such as the magnetic field) we can  encircle  the EP and so switch freely  between different magnon eigenmodes.\cite{PhysRevApplied034050} Such features are interesting for applications in magnonic logics, more sensitive  sensors for external magnetic changes, magnonic amplifiers, or applications to magnetic switching.  Recent studies concentrate on the  coupling between the magnetic waveguides via  the interlayer exchange (RKKY) interaction. Even though  RKKY coupled magnetic layers are experimentally feasible, the RKKY coupling is strongly sensitive to interface properties and requires (for our purpose) a high degree of control on the quality and thickness of the spacer layer.}

 A more general and versatile setup is offered by dipolarly coupled waveguides whose utility for magnonics has already been demonstrated.\cite{Wang2018,Wangq2020} 
 Here we propose and study the theoretical aspects of PT-symmetric dipolarly coupled waveguides. Using a combination of micromagnetic simulations and analytical theory, we present designs where two dipolarly coupled waveguides (WGs) deposited on the normal (heavy) metal show a PT-typical behavior when driving a current density in the substrate. 
 Contrasting with the RKKY case, the dipolar coupling strength is weaker, with the result that a relatively small electric current density is sufficient to drive the system between the PT-symmetry preserving and broken phases across the EP. In addition, for the dipolarly coupled WGs, we find a wave vector-dependent PT-symmetric behavior. In some ranges, where the dipolarly coupled magnon modes are degenerate, the electric current density at EP can approach zero. Varying the distance between the two WGs periodically, magnonic bands are formed \cite{Krawczyk_2014,Chumak_2017}. By driving the system to the vicinity of the EP, the magnonic amplitude is strongly enhanced at the Brillouin zone (BZ) boundary. This opens the possibility for having gaped spectrum with a wave vector-specific magnon-amplitude enhancement and increased sensitivity by adjusting the structuring period.

\section{Anti-parallel aligned magnetic stripes}

 \begin{figure}
	\includegraphics[width=0.82\textwidth]{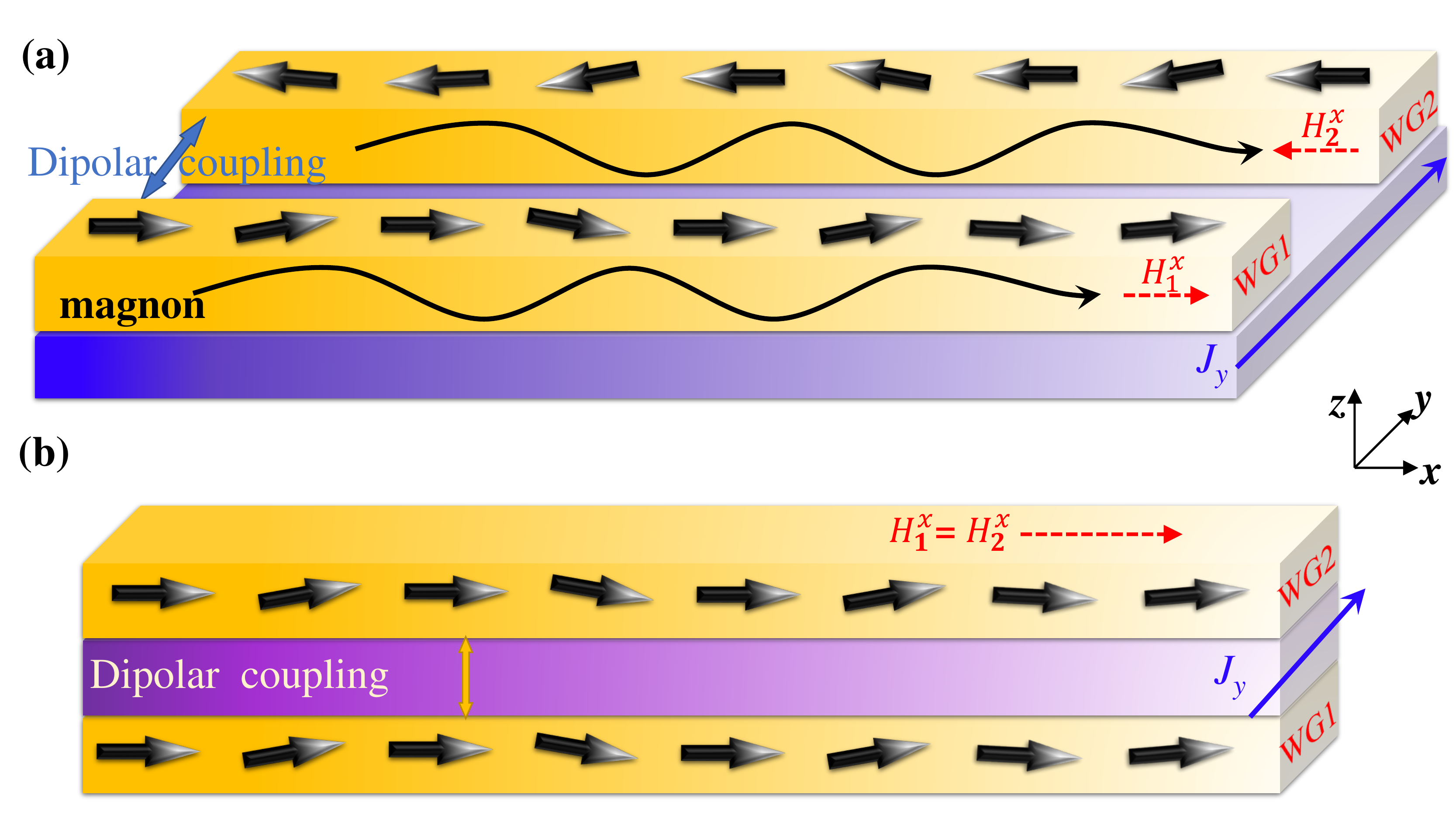}
	\caption{\label{model} (a) Schematics of two dipolarly coupled magnonic waveguides WG1 and WG2 (magnetized oppositely along $ x $ axis) and deposited on a heavy metal substrate (for example Pt). The magnons are launched locally at the left end ($ x = 0 $) in WG1 or WG2. Injecting in the Pt layer a charge current $ J_{y} $ results {in} spin-orbit torques (SOT) acting on both magnetic layers but with opposite directions. Therefore, the SOT damp or antidamp the magnons in WG1 or WG2 allowing so to realize a PT-symmetric system with the magnon propagation having particular features controllable by SOTs (or $ J_{y} $). { \bcorr The static fields $ H_1^x $ and $ H_2^x $ in WG1 and WG2 have equal magnitude and opposite directions. Such fields can be realized by attaching separated bias layers to the waveguides. (b) Schematic of two dipolarly coupled magnonic waveguides WG1 and WG2 (magnetized  along the  $ x $ axis). Injecting in the spacer Pt layer a charge current $ J_{\,\rm Pt} $ acts  with opposite SOTs on WG1 and WG2, meaning it damps one waveguide and antidamps the other. In this way, the PT-symmetry phases can be controlled and correspondingly the inherent feature of the magnon propagation.}}
\end{figure}

The proposed structure is shown in Fig.\,\ref{model}(a). The magnons are excited and propagate in two dipolarly coupled waveguides, WG1 and WG2. For triggering the magnons and manipulating the magnonic PT-symmetry, local magnetic fields, the substrate electric currents, and the interfacial spin Hall effects are exploited. 
 The latter leads to the spin-orbit torque (SOT) {$ \vec{T}_{p} = \gamma c_{J}\, \vec{m}_{p} \times \hat{\vec{x}} \times \vec{m}_{p} $} acting on the layer $ p = 1,2 $ with the magnetization unit vector field {$\vec{m}_{p} $}.
 SOT strength is $ c_{J} $ (with a sign depending on the charge current direction) , and $ \gamma $ is the gyromagnetic ratio. 
 If the magnetizations in the two WGs are opposite, the SOT results in losses and gains in the amplitudes of  the dipolarly coupled magnonic modes.

 Generally, the dipolarly coupled magnetization dynamics is governed by the Landau-Lifshitz-Gilbert (LLG) equations,\cite{Krivorotov228,Collet2016}
\begin{equation}
\displaystyle \frac{\partial \vec{m}_p}{\partial t} = - \gamma \,\vec{m}_p \times \vec{H}_{\,\rm{eff,}p}+ \alpha\, \vec{m}_p \times \frac{\partial \vec{m}_p}{\partial t} + \vec{T}_p.
\label{LLG}
\end{equation}
$ M_{\mr{S}} $ is the local saturation magnetization. 
$ \mu_0 $ is the vacuum permeability, and $ \alpha $ is the intrinsic Gilbert damping in the respective layer. The effective field {$ \vec{H}_{\,\rm{eff},p} = \frac{2 A_{\mathrm{ex}}}{\mu_0 M_{\mr{S}}} \nabla^2 \vec{m}_p + H^x_{p} \hat{\vec{x}} + \vec{H}_{\mathrm{demag}} $} consists of the internal exchange field (with an exchange constant $ A_{\mathrm{ex}} $), the external magnetic field $ H^x_{p} $, and the demagnetization field. The magnetic dipole-dipole interactions are included in the demagnetization field:
\begin{equation}
\displaystyle \vec{H}_{\mathrm{demag}}(\vec{r}) = -\frac{M_{\mr{S}}}{4 \pi} \nabla\int_{V'} \nabla'\frac{1}{|\vec{r} - \vec{r'}|} \cdot \vec{m}(\vec{r'})\,\dd V' .
\label{demag}
\end{equation}
We study the magnon dynamics by using linearized analytical models, supported and complemented with numerical simulations of Eq.\,(\ref{LLG}) that allow assessing the validity of the linear regime.

 The predictions are generic but for the sake of demonstration, we perform the simulations for 
magnetic layers made of Yttrium iron garnet (Y$_3$Fe$_2$(FeO$_4$)$_3$, YIG) corresponding to 
$ M_{\mr{S}} = 1.4 \times 10^5 \,$A/m, $ A_{\mathrm{ex}} = 3 \times 10^{-12} \,$J/m, and Gilbert damping $ \alpha = 0.001 $.
 The magnetic field along the axis of the waveguide ($ x $ axis) is $ H^x_1 = -H^x_2 = 10 \,$mT. {\bcorr The small bias field $ H_p^x $ can be generated via coupling fields from attaching bias layers.} The two stripes have the same extensions with a width of $ w = 100 \,$nm (along $ y $ axis), thickness $ h = 50 \,$nm (along $ z $ axis). The space between the WGs is $ \delta = 100 \,$nm wide. 

 \begin{figure}[!h]
	\includegraphics[width=0.72\textwidth]{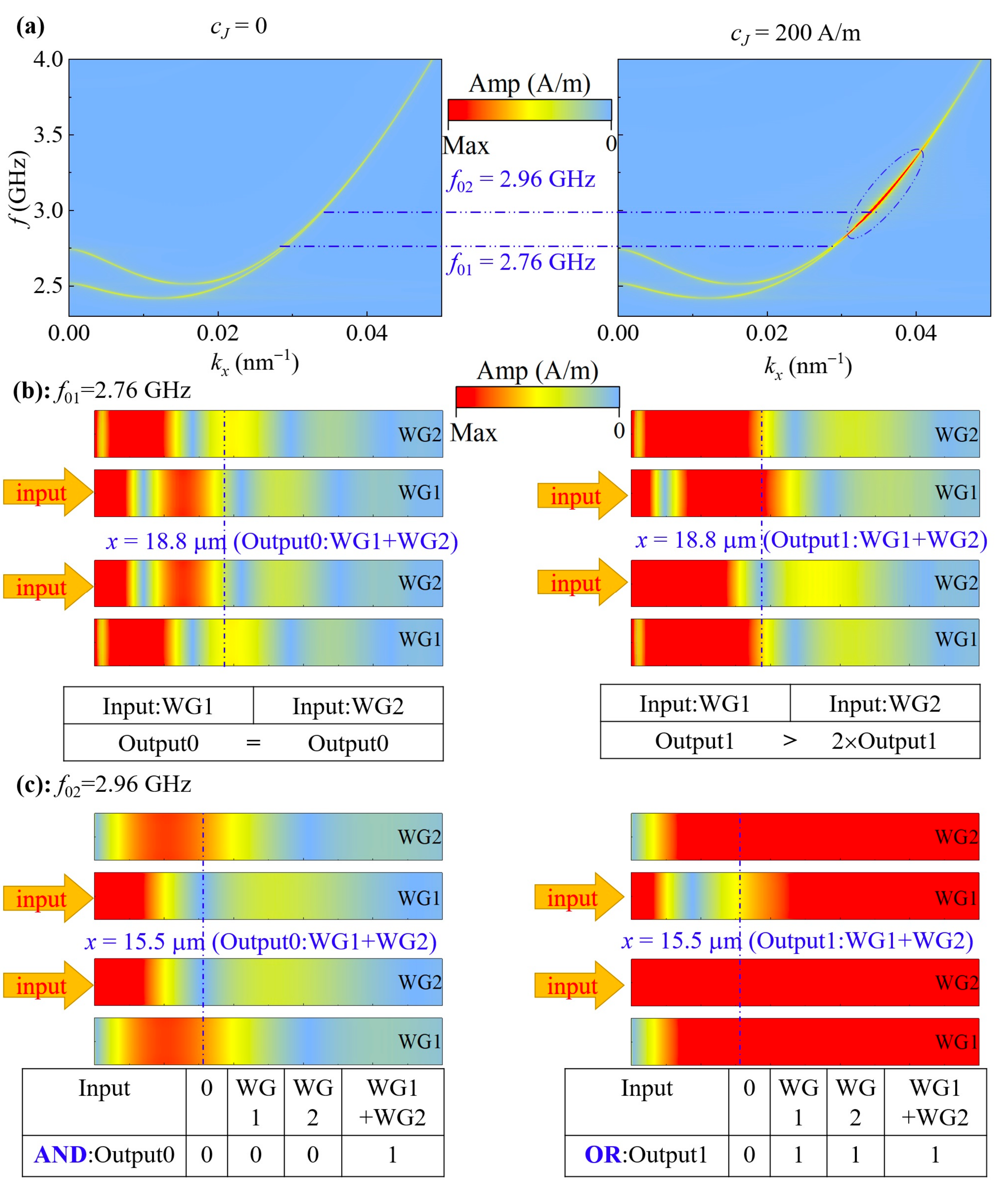}
	\caption{\label{propa1} \bcorr (a) Simulated SW dispersion curves (frequency $f$ vs. wave vector $k_x$) of the magnons (showing the optic and acoustic modes) of the dipolarly coupled waveguides. At the frequency (b) $ f = 2.76 $ GHz and (c) $ f = 2.96 $ GHz, the spatial profiles of the magnon amplitudes in WG1 and WG2. The injected SWs are excited at $ x = 0 $ in WG1 (i.e., Input:WG1) or WG2 (i.e., Input:WG2). The outputs are detected via the sum of SW amplitudes in WG1 and WG2 at (b) $ x = 18.8 {\rm \mu m} $ or (c) $ x = 15.5 {\rm \mu m} $. Here, left and right columns are for $ c_J = 0 $ and $ c_J = 200 \,$A/m, respectively.}
\end{figure}

To launch the magnons, we start from the initially anti-parallel magnetization ($ \vec{m}_1 = (1,0,0) $ and $ \vec{m}_2 = (-1,0,0) $) and relax to the stationary state. A periodic magnetic field pulse $ h(t) = h_a \vec{z} \sin(2 \pi f_H t) / (2 \pi f_H t) $ is then applied locally to WG1 at $ x = 0 $ which results in propagating magnons over a wide frequency range. Here, the excitation cutoff frequency is $ f_H = 25 $ GHz, and the field amplitude is $ h_a = 25 \,$mT. 
In the simulations, we analyze the magnetization fluctuations extracted from each cell $ M_z $ during 200 ns with a time step of 20 ps. To obtain the magnon dispersion relation, we use a two-dimensional fast Fourier transform $ M_z (k_x, f) = (1/N)\sum_{q=1}^{N_y}F_2[M_z(x,y_q,t)] $, where $ y_q $ is the $ q $th cell along the $ y $ axis, where $ N_y $ is the total number of cells along the $ y $ axis. The cell size adopted in the simulation is $ 20 \,\rm{nm} \times 20 \,\rm{nm} \times 50 \,\rm{nm} $. {\bcorr In the coupled waveguides with strong shape anisotropy, the parallel or anti-parallel configurations are both stable.} To realize the initial anti-parallel magnetizations, at first we drive the system to a state of parallel magnetizations in two magnetic layers using a high magnetic field {\bcorr (much larger than the amplitude of the small bias field $ H_p^x $)}. Then, a short perpendicular magnetic-field pulse triggers the flip between the parallel and anti-parallel states.\cite{Wang2018,PhysRevB87134419,Haldar2016}

The numerical magnon dispersion relation obtained for two dipolarly coupled waveguides is shown in Fig.\,\ref{propa1}(a). {\bcorr The dispersion curves exhibit similar features with backward volume SW modes, with the group velocity being negative in the low wave-vector range ($ |k_x| < 0.015 {\rm nm}^{-1} $), and the exchange term dominates the dipolar term in the larger $ k_x $ range.} The dipolar coupling between WG1 and WG2 leads to the formation of lower "acoustic" and higher "optic" modes. Due to the nature of dipolar coupling, in the low frequency range with large wave-length, the difference between the two magnon modes is more obvious, and declines with increasing frequency $ f $. At a certain $ f $, the two magnon modes with different wave vectors are excited simultaneously, and their interference leads to a periodic transfer of magnonic power between the WGs, as demonstrated by Fig.\,\ref{propa1}(b). The coupling length is larger for larger frequencies, where the wave vector difference between the two modes is smaller, see the left column of Fig.\,\ref{propa1}(a-c). {\bcorr Furthermore, the splitting between the two SW modes decreases with the spacing thickness between the two guides, as the dipolar coupling strength increases while the spacing thickness shrinks.}

For an analytical insight we assume the magnetization vectors to deviate slightly from their equilibrium states (long-wavelength SW limit), meaning {$ \vec{m}_p (\vec{r},t) = \vec{m}_{0,p} + \vec{m}_{s,p} \ee^{\ii \left(\vec{k}_s \cdot \vec{r} - \omega t \right)} $}. Here, $ \vec{m}_{0,1} = -\vec{m}_{0,2} = \hat{\vec{x}} $ is the static anti-parallel equilibrium magnetization, and the small deviation from equilibrium is $ \vec{m}_{s,p} = (0, \delta m_{y,p},\delta m_{z,p}) $ with $ \delta m_{y(z),p} \ll 1 $. The wave vector $ \vec{k}_s $ is the sum of the in-plane wave vectors $ \vec{k} = k_x \hat{\vec{x}} + k_y \hat{\vec{y}} $. In Fourier space, the linearized LLG equation with SOT yields the following expression for the SWs \cite{Wangxinc2020,Wang2018,Verba2012}
 {
\begin{equation}
\displaystyle -\ii \omega \vec{m}_{s,p} = \vec{m}_{0,p} \times \sum_{q} \hat{\Omega}_{pq}\cdot m_{s,q} - \vec{m}_{0,p} \times (\ii \alpha \omega \vec{m}_{s,p} - \omega_J \hat{\vec{x}} \times \vec{m}_{s,p}).
\label{swequation}
\end{equation}}
 Here, $ \omega_J = \gamma c_J $, and the tensor $ \hat{\Omega}_{pq} $ has the form {
\begin{equation}
\displaystyle \hat{\Omega}_{pq} = \Omega_{0} \delta_{pq}\hat{I} + \omega_{\rm{M}} \hat{F}(d_{|p-q|}).
\label{tensorO}
\end{equation}}
We defined {$ \Omega_{0} = \gamma H_1^x + \frac{2 \gamma A_{\,\rm{ex}} k^2}{\mu_0 M_{\mr{S}}} $}, and $ \omega_{\rm{M}} = \gamma M_{\mr{S}} $. The wave vector $ k $ is equal to $ \sqrt{k_x^2 + k_y^2} $, the distance between {the two consecutive WGs is $ d_{1} = w + \delta $}. The dynamic magnetodipolar interaction is described by the tensor $ \hat{\vec{F}} $: 
\begin{equation}
\begin{aligned}
\displaystyle \hat{\vec{F}}(d_p) &= \int\hat{\vec{N}}_k \ee^{\ii k_y d_p} \frac{\dd k_y}{2\pi},\\
\hat{N}^{\alpha\beta}_k &= \frac{\sigma_s^2(k_y)}{w_s}\int D_p(k_z) D_q^*(k_z)\frac{k_\alpha k_\beta}{k^2} \frac{\dd k_z}{2\pi}.
\label{tensorF}
\end{aligned}
\end{equation}
The "shape amplitude" {$ D_p(k_z) = \int_{-h/2}^{h/2} \ee^{-\ii k_z z} \,\dd z = h \,\mr{sinc}\!\left(h k_z / 2\right)$} describes the influence of the finite thickness $ h $ of the thin WG. Here, $ \sigma_s = \int_{-w/2}^{w/2} m(y)\, \ee^{-\ii k_y y} \,\dd y $ is the Fourier transform of the SW profile $ m(y) $ across the width direction, and the effective width {$ w_s = \int_{-w/2}^{w/2} m^2(y) \,\dd y$}. For a narrow WG we may assume $ m(y) \sim \cos(\kappa y) $ for $ \kappa = 0 $, in which case the effective width is equal to the WG width and $\sigma_s = w \, \mr{sinc}\!\left(w k_y / 2\right)$.
Introducing $ \psi_p^\pm = \delta m_{y,p} \pm \ii \delta m_{z,p} $, we obtain the determining equation $ \omega \vec{\psi} = \hat{H} \vec{\psi} $ for the eigenmodes $ \vec{\psi} = (\psi_1^+, \psi_1^-, \psi_2^+, \psi_2^-) $ at the SWs eigenfrequencies $\omega$. Inspecting Eq.~(\ref{swequation}), we infer the structure of the $ 4 \times 4 $ operator $ \hat{H} $ as
\begin{align}\label{haml}
	\hat{H}=
	\left(
	\begin{array}{cccc}
		-\tilde{\omega}^--\ii\omega _J^- & \Delta \omega _0^- & -\bar{\omega }_1^- & \Delta \omega _1^- \\
		-\Delta \omega _0^+ & \tilde{\omega}^+-\ii\omega _J^+ & -\Delta \omega _1^+ & \bar{\omega }_1^+ \\
		\bar{\omega }_1^+ & -\Delta \omega _1^+ & \tilde{\omega}^++\ii\omega _J^+ & -\Delta \omega _0^+ \\
		\Delta \omega _1^- & -\bar{\omega }_1^- & \Delta \omega _0^- & -\tilde{\omega}^-+\ii\omega _J^- \\
	\end{array}
	\right)\, .
\end{align}
We introduced the abbreviation $a/(1\pm \ii \alpha)\equiv a^\pm$ for arbitrary $a$, and 
 $\Delta\omega_{j} = \omega^z_j - \omega^y_j$ is the difference of the demagnetizing field for the $y$ and $z$ component, $\bar{\omega}_{j} = \omega^y_j + \omega^z_j$ is the sum of the demagnetizing field for the $y$ and $z$ component, and $\tilde{\omega} = \Omega_{0} + \bar{\omega}_0$, where $\omega^y_j = \frac{\gamma M_{\mr{S}}}{2} F^{yy}(d_j)$ and $\omega^z_j = \frac{\gamma M_{\mr{S}}}{2} F^{zz}(d_j)$. $d_1 = w + \delta$ is the distance between WG1 and WG2. The waveguide interacting with its own {\bcorr dynamic demagnetizing field} is captured for $j = 0$ ($ d_0 = 0 $).

 \begin{figure}
	\includegraphics[width=0.72\textwidth]{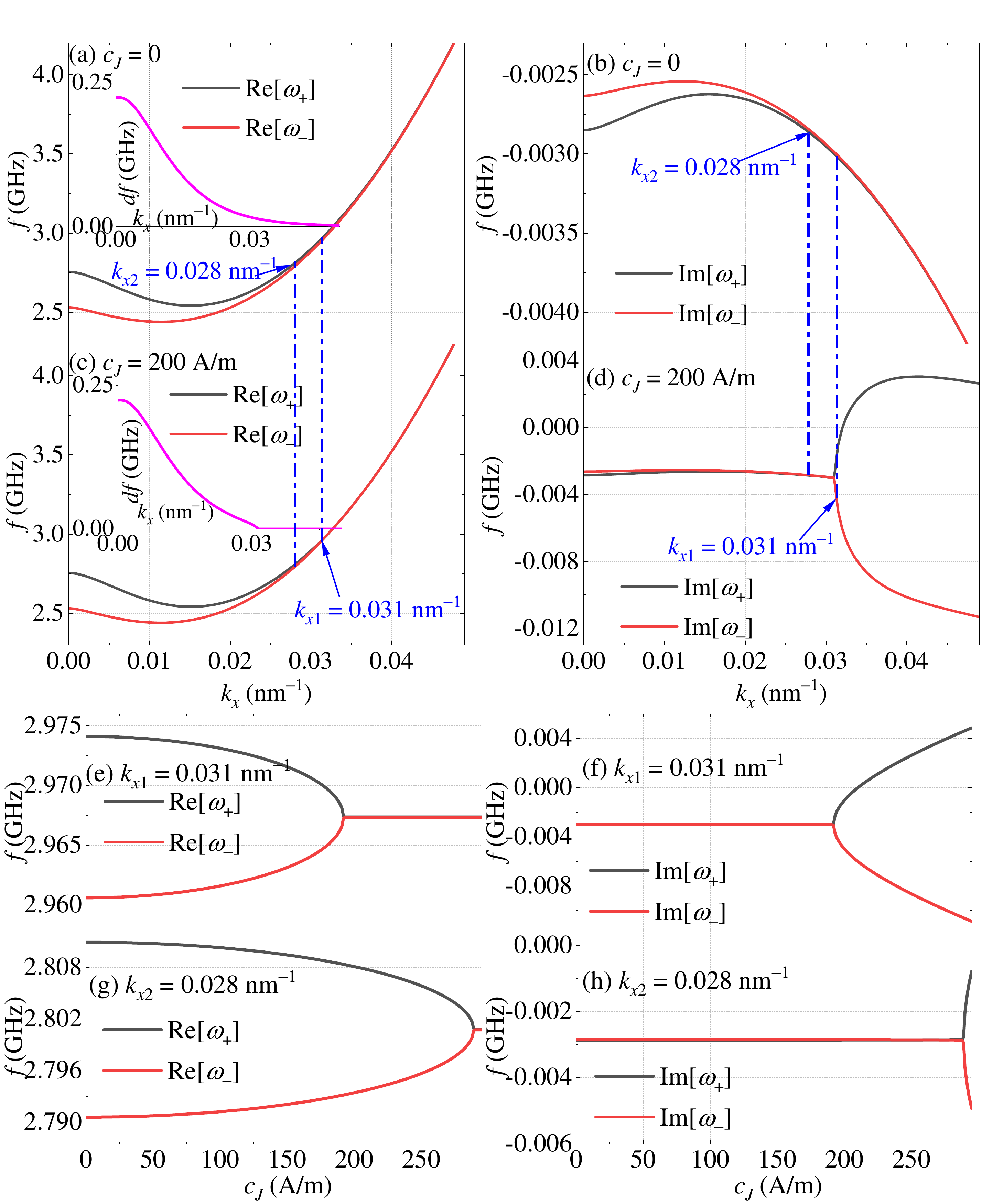}
	\caption{\label{ana1} \bcorr (a-d) The analytical linearized model yields the real (dispersion) and the imaginary parts of two magnon eigenfrequencies $ f = \omega/ (2 \pi) $ in the dipolarly coupled waveguides. Eigenfrequencies are obtained from the two positive eigenvalues of the matrix  given by Eq. (\ref{haml}). The results are shown for $ c_J = 0 $ (a-b) and $ c_J = 200 \,$A/m (c-d). The insets in (a) and (c) show $ df = {\rm Re}[\omega_+ - \omega_-]/ (2 \pi) $. (e-h): The results of the analytical model for the real and the imaginary parts of the two magnon eigenfrequencies $ f = \omega/ (2 \pi) $, as we scan $ c_J $ at the wave vector $ k_x = 0.031 \,\rm{nm}^{-1} $ and $ 0.028 \,\rm{nm}^{-1} $.}
\end{figure}

Without SOT ($ \omega_J = 0 $), we obtain four eigenfrequencies $ \omega $ for $ \hat{H} $. Two positive frequencies correspond to right-handed precessions around the stable state, while the other two negative frequencies are associated with left-handed precessions. Applying a finite SOT with $ c_J = 200 \,$A/m, the real parts of the two magnon modes turn identical for larger $ k_x $ (Fig. \,\ref{ana1}(a-b)), and the two imaginary parts are obviously separated (see Fig.\,\ref{ana1}(c-d)). Considering the dependence of the eigenfrequencies on $ c_J $ at different $ k_x $ {(Fig. \ref{ana1}(e-h))}, for very small $ c_J $ the optical and the acoustic modes are still separated, and the two modes merge at a point (identified as the exceptional point, EP) when increasing $ c_J $. Beyond the EP, the real parts of the two modes collapse at the same value and the imaginary parts bifurcate. This is a scenario known for EP in PT-symmetric systems. An interesting feature here is that the amplitude of $ c_J $ at the EP is dependent on $ k_x $. For smaller $ k_x $, the value of $ c_J $ for the occurrence of EP is larger. This explains why the two imaginary parts bifurcate at higher $ k_x $ for $ c_J = 200 \,$A/m in Fig.\,\ref{ana1}(c-d), as the magnons for low $ k_x $ are still below the EP, while they have surpassed the EP for higher $ k_x $. {\bcorr Despite the relative simplicity and approximations, the analytical results agree well with those of micromagnetic simulations in Fig. \ref{propa1}(a-b). As well, in the simulations, the SW excitation is strongly enhanced at higher $ k_x $ when $ c_J = 200 \,$A/m, which is in line with the altered imaginary parts in the analytical results in this regime. But there is still a slight quantitative difference between the simulations and the analytical results due to the approximations made. For example, when $ c_J = 200 \,$A/m there is a slight difference in the $ k_x $ range beyond the EP.}

The changes in the two magnon modes affect the interference and the energy transfer between the WGs, as evidenced by the simulation results in Fig.\,\ref{propa1}(b-c). {\bcorr For $ c_J=0 $, meaning  the conventional case, at a certain frequency, the superposition of the two modes (symmetric and antisymmetric) leads to a magnon periodic transfer between WG1 and WG2. The periodic transfer length $ L $ is directly related to the wave vector difference between the two modes $ L = \pi /|k_{x,+} - k_{x,-}| $.\cite{Wang2018} For a larger frequency, a smaller $ |k_{x,+} - k_{x,-}| $ leads to a larger transfer length (Fig. \ref{propa1}(b-c) when $ c_J = 0 $). Besides, by changing from the anti-parallel magnetization orientation  to  the parallel orientation, the coupling strength and $ |k_{x,+} - k_{x,-}| $ become  weaker/smaller (not shown), resulting in a larger transfer length $ L $ of the output power. Here, the magnon propagation is reciprocal. For example, in the left column of Fig. \ref{propa1}(b-c), when switching the excitation from WG1 to WG2, the SW profiles in the two waveguides are also switched to each other.}
	
{\bcorr When $ c_J $ is applied in the range below the EP (for example, $ c_J = 200 \,$A/m and $ f = 2.76 $ GHz in Fig. \ref{propa1} (b)), in the unbroken PT-symmetric regime there are still two modes, but the two modes are not orthogonal anymore \cite{Wangxinc2020,Ruter2010}. Thus, exchanging the input between WG1 and WG2 results in different SW profiles, i.e., the magnon propagation becomes non-reciprocal. Increasing the charge current $ c_J $ leads to a smaller $ |k_{x,+} - k_{x,-}| $, meaning  a larger transfer length $ L $. For the case above the EP (for example, $ c_J = 200 \,$A/m and $ f = 2.96 $ GHz in Fig. \ref{propa1} (c)), two modes coalesce and the SW can simultaneously propagate in both guides. Here, one can still find the non-reciprocal feature, where the WG2 with gain always has a larger SW amplitude when the input is exchanged.} 

{\bcorr The non-reciprocal propagation in the electric current-induced PT-symmetric system points to potential applications in reconfigurable diode and logic devices. In Fig. \ref{propa1}(b) with frequency $ f = 2.76 $ GHz, when $ c_J = 0 $ two equal outputs are obtained by switching the input (SW injection) between WG1 and WG2 at $ f = 2.76 $ GHz. Turning on the electric current ($ c_J = 200 $ A/m), the SW output for "Input:WG1" becomes twice the out of "Input:WG2". Such a non-reciprocal result clearly points to a realization of a reconfigurable SW diode. The input in WG1 represents an "on" output with a larger amplitude, and input in WG2 signals an "off" state with an obviously smaller amplitude. Considering Fig. \ref{propa1}(c) with frequency $ f = 2.96 $ GHz, when $ c_J = 0 $  and the excitation is  in both WG1 and WG2, only the symmetric SW mode is excited and the periodic propagation is off, resulting in the output with larger amplitude (logic "1") twice the output (logic "0") from other inputs (only in WG1 or WG2). This feature indicates a logic "AND" operation. When $ c_J = 200 $ A/m, the excitation is enhanced in both guides and the periodic transfer is also off above the EP. Here, larger SW amplitudes are achieved for input in WG1, WG2, or both waveguides, i.e., a logic "OR" operation, indicating an electrically reconfigurable SW logic device operated via the enhanced SW propagation above EP.}

{\bcorr The above analyses are only for anti-parallel WG1 and WG2. For parallel configuration, there is still splitting between the optic and acoustic modes. Compared with the anti-parallel case, the mode splitting becomes slightly smaller under parallel configuration (not shown). Applying SOT on the parallel case can enhance or weaken the damping of coupled modes, and in such cases, one can not achieve the PT-symmetry-phase change and EP. To realize the above effect in parallel waveguides, we provide a structure of Fig. \ref{model}(b), as discussed in the following section.} 

\section{Waveguides with parallel magnetizations}

 \begin{figure}
	\includegraphics[width=0.92\textwidth]{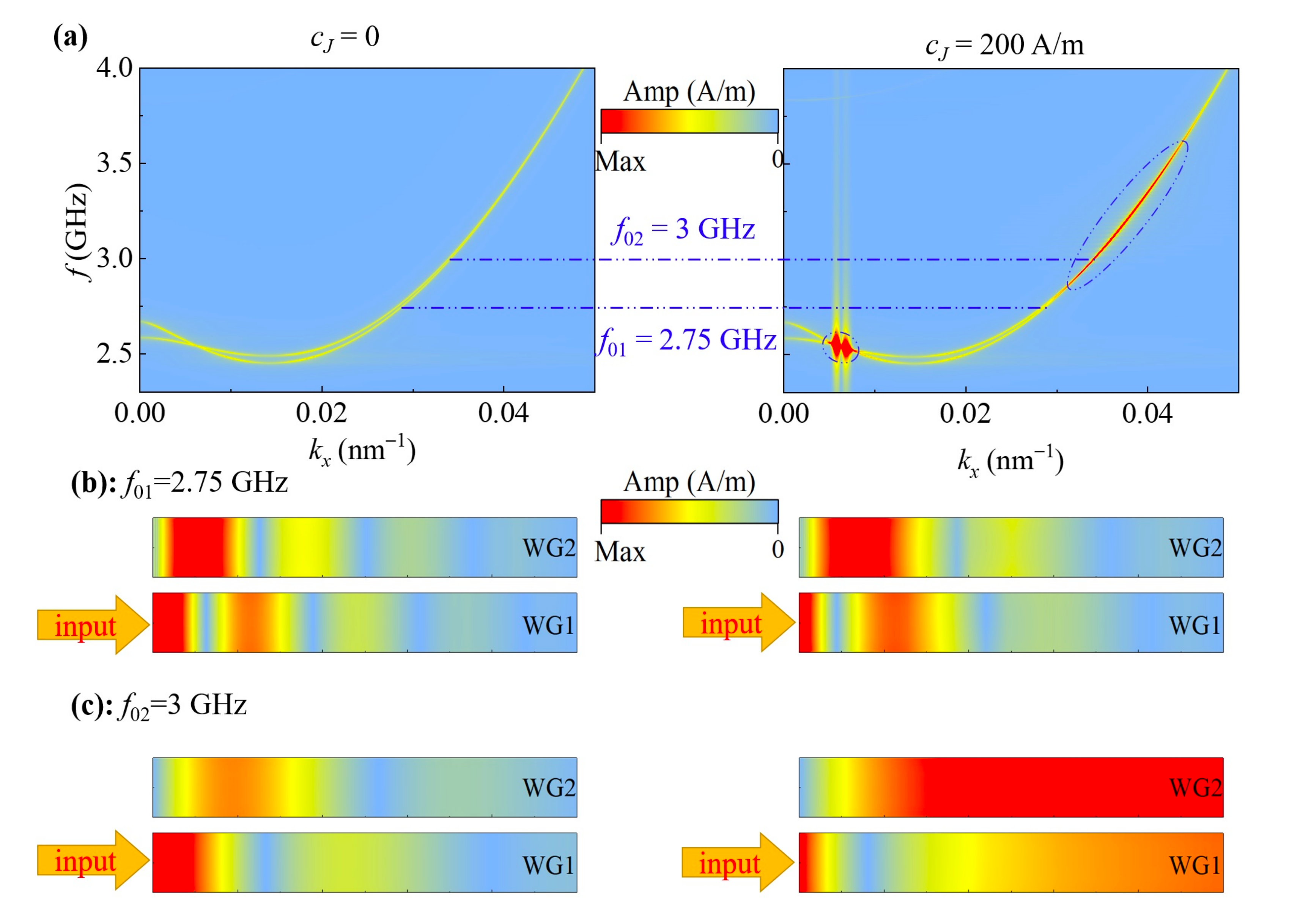}
	\caption{\label{propaga2} \bcorr Under the structure of Fig.\,\ref{model}(b), (a) the dispersion curves of two collective magnon modes (optic and acoustic modes) of coupled waveguides. (b-c) The simulated spatial profiles of the propagating SW amplitude at (b) $ f = 2.75 $ GHz and (c) $ f = 3 $ GHz. Here, the SWs are excited at $ x = 0 $ in WG1, and left and right columns are for $ c_J = 0 $ and $ c_J = 200 \,$A/m, respectively.}
\end{figure}

 \begin{figure}
	\includegraphics[width=0.72\textwidth]{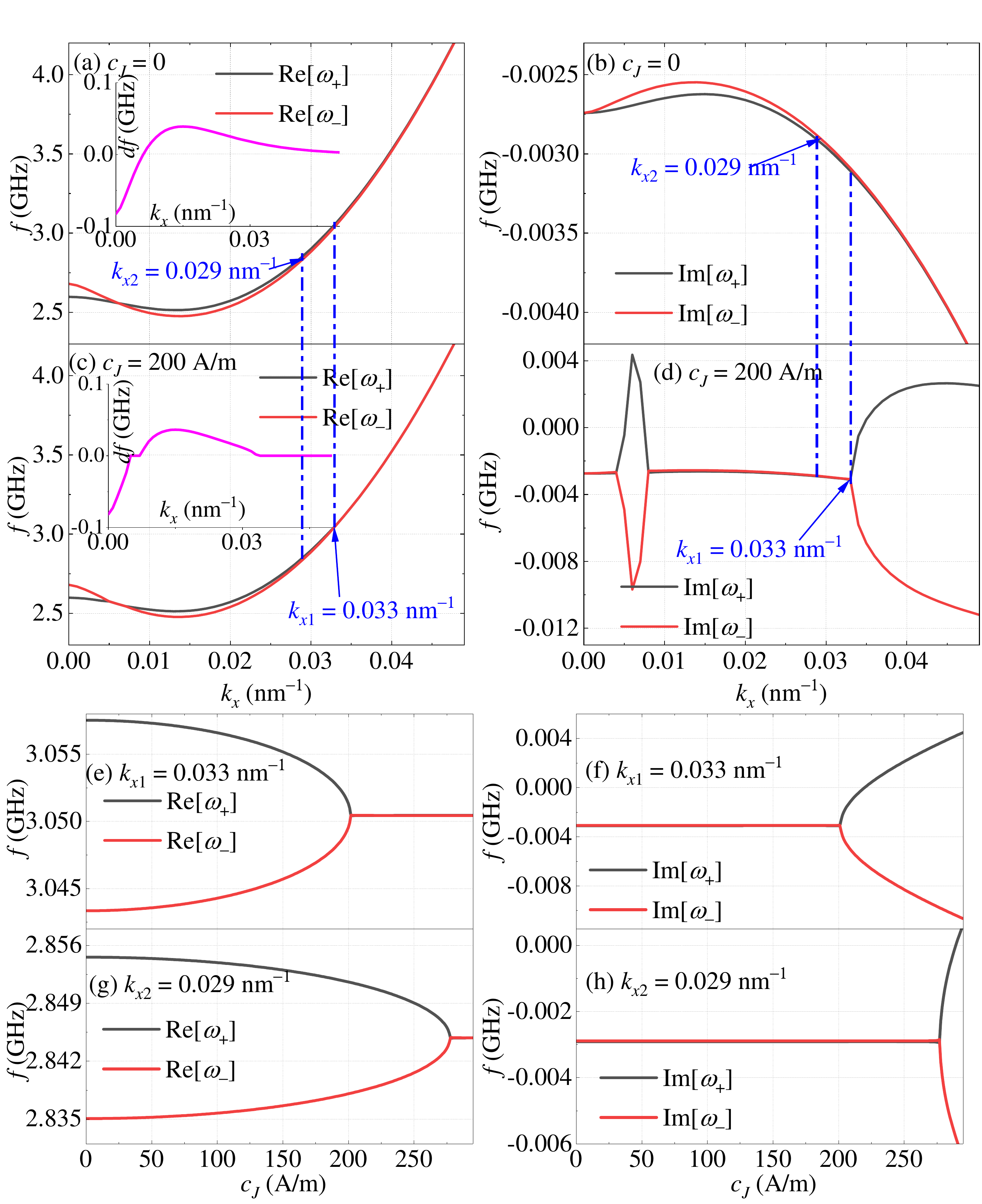}
	\caption{\label{ana2} \bcorr For the structure of Fig.\,\ref{model}(b), numerically solving the analytical model delivers  the real (dispersion) and the imaginary parts of the two magnon eigenfrequencies $ f = \omega/ (2 \pi) $ in the dipolarly coupled waveguides for $ c_J = 0 $ (a-b) and $ c_J = 200 \,$A/m (c-d). Here, the eigenfrequencies are obtained by the two positive eigenvalues of the magnon Hamiltonian (similar to Eq. (\ref{haml})). The insets in (a) and (c) show $ df = {\rm Re}[\omega_+ - \omega_-]/ (2 \pi) $. (e-h):  Real and imaginary parts of two magnon eigenfrequencies $ f = \omega/ (2 \pi) $ as we scan $ c_J $ at the wave vector $ k_x = 0.033 \,\rm{nm}^{-1} $ and $ 0.029 \,\rm{nm}^{-1} $.}
\end{figure}

It is also possible to realize dipolarly coupled, PT-symmetric WGs with parallel magnetizations, as illustrated in Fig.\,\ref{model}(b).
 The spacer induced SOT reads {$ \vec{T}_{p} = (-1)^p \gamma c_J \vec{m}_{p} \times \hat{\vec{x}} \times \vec{m}_{p} $}. 
 The simulation results for the dispersion relation are shown in Fig.\,\ref{propaga2}(a) for a structure
with a width $ w = 100 \,$nm (along $ y $ axis), thickness $ h = 50 \,$nm (along $ z $ axis), and a spacer thickness of $ \delta = 100 \,$nm.
The dispersion curves divert when the two modes cross at the low frequency range (Fig.\,\ref{propaga2}(a)). For larger frequencies, the curves are similar to Fig.\,\ref{propa1}(a). The interference-induced power transfer between WG1 and WG2 is obvious at $ f = 2.75 $ GHz (Fig.\,\ref{propaga2}(b)).

Following similar derivation steps, an analytical analysis is performed.
The SW dispersion relation is presented in Fig.\,\ref{ana2}.
% which agrees well with the respective numerically evaluated coupled dispersion relation curve.
 For $ c_J = 200 \,$A/m, the real parts and the imaginary parts are shown in Fig.\,\ref{ana2}(c-d). Still, we observe that the real parts collapse and the imaginary parts bifurcate at higher $ k_x $. In contrast, for lower $ k_x $, the two modes cross even without SOT ($ c_J = 0 $). 
 For $ c_J = 200 \,$A/m the range with identical real parts of frequencies is enlarged, and in this range the imaginary parts are strongly separated. We note the cross modes decrease the required current density of EP to be around 0, much smaller than the current density needed for the auto-oscillation. Furthermore, by scanning $ c_J $, we still find a $ k_x$ dependence of the EP, as evidenced by Fig.\,\ref{ana2}.

 \begin{figure}
	\includegraphics[width=0.82\textwidth]{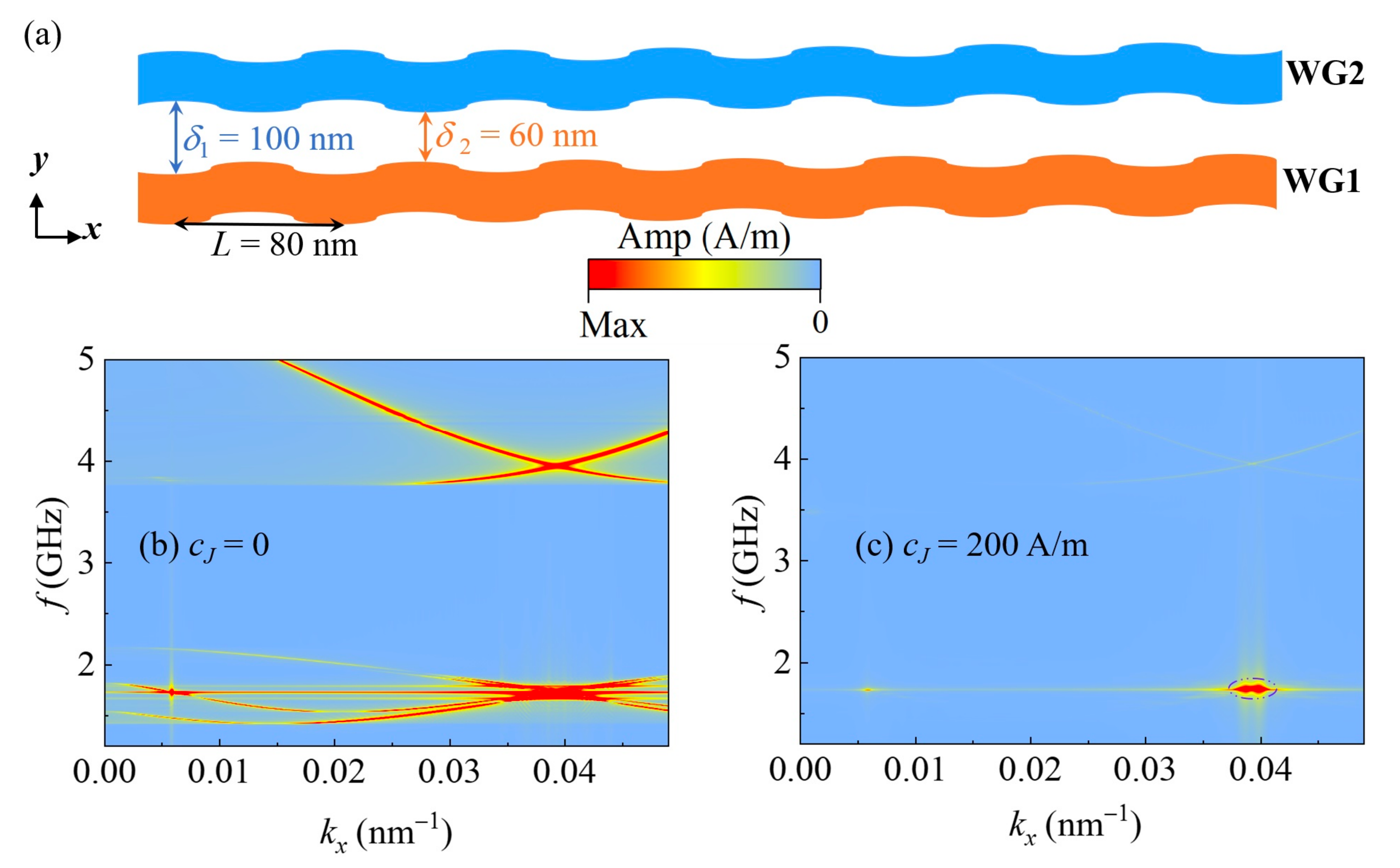}
	\caption{\label{crystal} (a) Schematic of periodic waveguides. The distance varies periodically between $ D_1 = 100 \,$nm and $ D_2 = 60 \,$nm, and period length $ L = 80 \,$nm. {\bcorr For such a structure, the Brillouin Zone (BZ) boundary is at $ k_x = \pi/L = 0.04 \,$nm$ ^{-1} $.} (b-c) Dispersion curves in the periodic waveguides for $ c_J = 0 $(b) and $ c_J = 200 \,$A/m (c).}
\end{figure}

To endorse the above analysis, we simulate full numerically the SW propagation. As $ c_J $ approaches the EP, the change in the SW beating is shown by Fig.\,\ref{propaga2}. A change in the imaginary parts at larger $ k_x $ and near the crossing point is evidenced by the SW excitation in the right column of Fig.\,\ref{propaga2}(a), where the SW amplitude is strongly enhanced in the range above EP. Increasing the number of WGs does not alter qualitatively the features discussed above. An example of four WGs and some physical P and T operators \cite{Mostafazadeh2003} are presented in the supplementary materials.\cite{supp}

\section{PT-symmetric magnonic crystal}

\begin{figure}
	\includegraphics[width=0.7\textwidth]{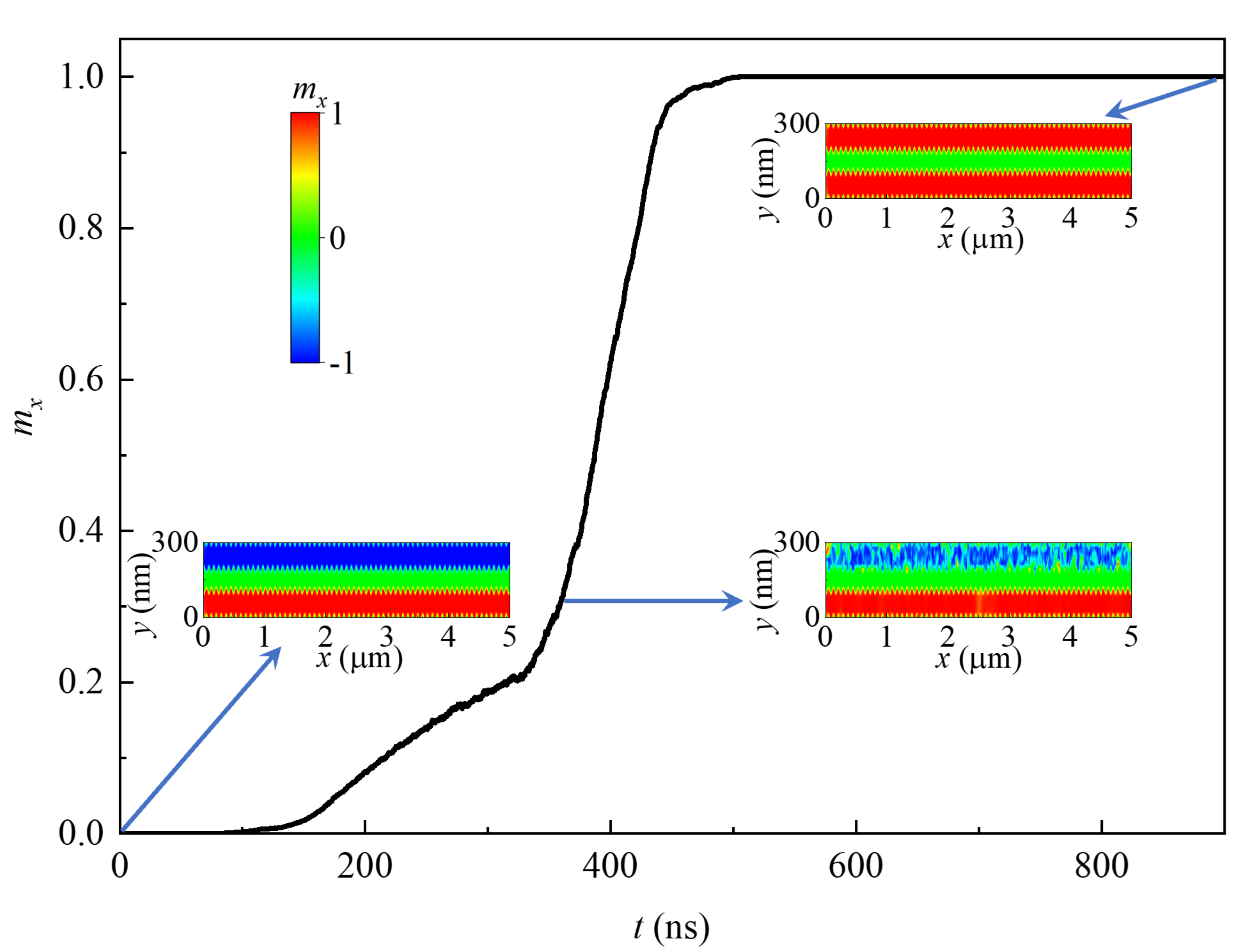}
	\caption{\label{reversetime} Time dependence of total $ m_x $ of WG1 and WG2 for $ c_J = 350 \,$A/m. Spatial profiles of $ m_x $ at different time points are shown as insets.}
\end{figure}

\begin{figure}
	\includegraphics[width=0.7\textwidth]{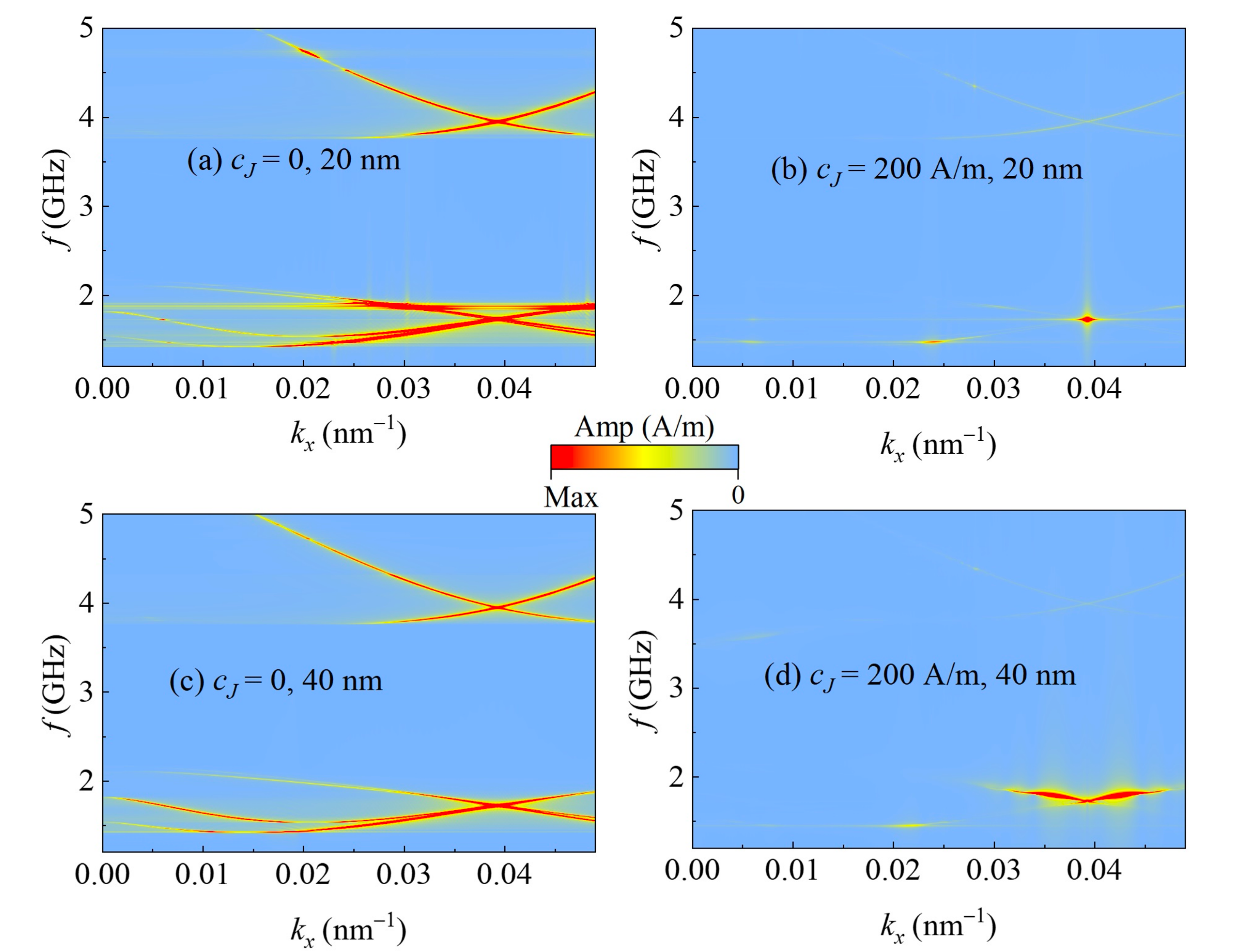}
	\caption{\label{crystalmove} Shifting the WG1 along $ x $ direction by the distance (a-b) 20 nm or (c-d) 40 nm, simulated SW dispersion relation of the coupled waveguides for $ c_J = 0 $ and $ c_J = 200 \,$A/m. }
\end{figure}

{\bcorr In the preceding sections, our results demonstrate a feasible way for the realization of bridging the PT-symmetry phases  when the two modes approach each other. In a magnonic crystal, the SW dispersion relation is folded at the Brillouin zone boundary, where different magnon modes are merged, providing an alternative solution to easily drive the broken PT-symmetry phase and manipulate the SW propagation.} The importance of having a gaped spectrum with controllable dispersion is well-documented in various areas such as {\bcorr electronics, photonics, and magnonics.\cite{Vlasov2005,Akahane2003,Ganainy2018,Konotop16,PhysRevB.84.205325,Chen2013,Pan2020,Chumak_2017,PhysRevLett.104.207205,PhysRevB.95.134433} In particular, the transport properties are intimately related to the dispersion characteristics of the underlying structure, and determine for an electronic system for example, whether we have a band insulator or a metal. Electronic correlation introduces nonlinearities and multiple scattering between the otherwise independent  effective single particles. Likewise, we expect for a magnonic system that introducing PT-symmetry would allow  amplifying non-linear effects at a particular point in the dispersion which could be of relevance to control multi-magnon scattering/generation in a certain area at the BZ}. As a demonstration we consider the periodic magnetic  structure (illustrated by Fig.\,\ref{crystal}(a)). The distance between WG1 and WG2 varies periodically. We note that in principle, comb-type magnetic stripes have already been fabricated.\cite{Lewis2010}
The periodic modulations are reflected in correspondingly  spatial changes in the internal effective fields that act on the SW and generate so a magnonic crystal.{\bcorr \cite{Kruglyak2010,Tkachenko2012,PhysRevLett.102.127202, Dikjap2014,Gubbiotti2018}} 
As shown in Fig.\,\ref{crystal}(b), the magnon band is folded at the Brillouin Zone (BZ) boundary ($ k_x = 0.04 \,$nm$ ^{-1} $). A clear band gap is identified between 2\,GHz and 4\,GHz, where the magnon cannot propagate. {\bcorr We note, that the band gap at the BZ boundary approaches 0 in our model,  similar to other magnonic crystals under particular geometric or magnetic parameters.\cite{SangKoog2009, PhysRevB.95.134433} } Applying a charge-current density to the waveguides results, as above, in opposite SOTs the two magnetic stripes. The whole structure constitutes a PT-symmetric magnonic crystal. From the preceding discussion we expect the PT-symmetry induced changes to be most prominent when the coupled bands cross. Here, the crossing occurs at the BZ boundary, and the exciting magnon amplitude is strongly enhanced, as demonstrated by the red color in the dispersion curve (Fig.\,\ref{crystal}(c)). Increasing $ c_J $ further amplifies the magnon excitation and finally leads to the magnetization reversal, as demonstrated by the time dependent $ m_x $ in Fig.\,\ref{reversetime} ($ c_J = 350 \,$A/m). The position of the wave vector dependent enhancement of the magnetic excitations is related to the spatial period of the modulations of the magnetic stripes and can be engineered desirably, allowing so to study and use non-linear SWs at a controllable wave vector. As shown in Fig.\,\ref{crystalmove}, we shift the WG1 in $ +x $ direction by $20\,$nm or $40\,$nm, the region with enhanced magnonic excitation is varied. {\bcorr For the displacement of $20\,$nm, a strongly localized magnon is excited near $ f = 1.88\,$GHz, causing the spectrum leakage (flat curve) in our calculation. As the Brillouin zone boundary is controllable via the periodic parameter, one can use the electric current to selectively enhance the SW at the BZ boundary and control the emitting SW from the magnonic crystal structure via the above effect.}

\section{Conclusions}
We presented proposals for the designs of magnetic structures that serve for magnonic signal steering and amplification based on dipolarly coupled waveguides with a loss-gain mechanism induced by a charge current in a metallic substrate hosting a strong spin orbit coupling.
 The magnonic signal can be steered and amplified by changing the current density and can thus be useful for spintronics. {\bcorr As a demonstration, we showed how the proposal can be employed as electrically reconfigurable magnonic diodes and for realizing logic operations. A particular advantage is that a relatively small current density is sufficient for signal control. Especially, when the magnonic bands approach each other, the required  current density needed to approach the PT-symmetry phase transition decreases. The band dispersion can be modulated and gaped by using a periodic PT-symmetric structure. In addition to potential use in magnonic devices, the proposed magnonic PT-symmetric crystal can serve to demonstrate the interplay between symmetry, damping, and nonlinear effects on the band structure of waves scattered from a periodic potential.} 

\section{ACKNOWLEDGMENTS}
This work was supported by the DFG through SFB TRR227, and Project Nr. 465098690, the National Natural Science
Foundation of China (Grants No. 12174452, No. 12074437, No. 11704415, No. 11674400, and No. 52073308), and the Natural Science Foundation of Hunan Province of China (Grants No. 2022JJ20050, No. 2021JJ30784, and No. 2020JJ4104).

\bibliographystyle{apsrev4-1}
\bibliography{PT-dipolar}

\end{document}